\documentclass[
aip
 amsmath,amssymb,
 reprint,%
]{revtex4-1}

\usepackage{graphicx}
\usepackage{dcolumn}
\usepackage{bm}

\usepackage[utf8]{inputenc}
\usepackage[T1]{fontenc}
\usepackage{mathptmx}
\usepackage{etoolbox}

\makeatletter
\def\@email#1#2{%
 \endgroup
 \patchcmd{\titleblock@produce}
  {\frontmatter@RRAPformat}
  {\frontmatter@RRAPformat{\produce@RRAP{*#1\href{mailto:#2}{#2}}}\frontmatter@RRAPformat}
  {}{}
}%
\makeatother
\begin{document}

\preprint{AIP/123-QED}

\title[Compact atom source using fiber-based pulsed laser ablation]{Compact atom source using fiber-based pulsed laser ablation}
\author{A. Osada}
\email{alto@g.ecc.u-tokyo.ac.jp}
 \affiliation{Komaba Institute for Science (KIS), The University of Tokyo, Meguro-ku, Tokyo, 153-8902, Japan}
 \affiliation{PRESTO, Japan Science and Technology Agency, Kawaguchi-shi, Saitama}

\author{R. Tamaki}%
\affiliation{Komaba Institute for Science (KIS), The University of Tokyo, Meguro-ku, Tokyo, 153-8902, Japan}

\author{W. Lin}%
\affiliation{Komaba Institute for Science (KIS), The University of Tokyo, Meguro-ku, Tokyo, 153-8902, Japan}

\author{I. Nakamura}
\affiliation{Komaba Institute for Science (KIS), The University of Tokyo, Meguro-ku, Tokyo, 153-8902, Japan}

\author{A. Noguchi}
 \affiliation{Komaba Institute for Science (KIS), The University of Tokyo, Meguro-ku, Tokyo, 153-8902, Japan}
 \affiliation{PRESTO, Japan Science and Technology Agency, Kawaguchi-shi, Saitama}
 \affiliation{RIKEN Center for Quantum Computing (RQC), Wako, Saitama 351--0198, Japan}
 \affiliation{Inamori Research Institute for Science (InaRIS), Kyoto-shi, Kyoto 600-8411, Japan }

\date{\today}

\begin{abstract}
We designed, demonstrated, and characterized an atom source based on fiber-based pulsed laser ablation.  By using commercially available miniature lens system for focusing nanosecond pulsed laser of up to 225~$\mu$J delivered through a multimode fiber of 105~$\mu$m core, we successfully ablate a SrTiO$_3$ target and generate a jet of neutral strontium atoms, though our method can be applied to other transparent ablation targets containing materials under concern.  Our device endures 6\,000 cycles of pulse delivery and irradiation without noticeable damage on the fiber facets and lenses.  The generated strontium beam is characterized with spectroscopic method and is revealed to exhibit the transverse temperature of 800~K and longitudinal velocity of 2\,300~m/s, which are typical of pulsed-laser-ablation-based atom source. The number of atoms generated by a single ablation pulse is estimated to be $2\times 10^5$. Our device provides a compact, cryo-compatible fiber-pigtailed atom source with minimized device footprints and reduced complexity of vacuum systems to further promote the developments of cold-atom experiments. It may also find interesting applications in atomic and molecular sciences.
\end{abstract}

\maketitle



Cold-atom quantum technologies has been developed as a platform of various scientific investigations such as quantum computing~\cite{Cirac1995-bz,Brown2016-me}, quantum networking~\cite{Connell2017-yg, Matthias_Keller_Birgit_Lange_Kazuhiro_Hayasaka_Wolfgang_Lange_Herbert_Walther2004-ib, Schupp2021-os}, quantum sensing~\cite{Dehmelt1982-hh,Burt2021-no,Brewer2019-jh} and quantum simulation~\cite{Georgescu2014, Greiner2002-bl}, and their progresses has gradually brought themselves to a field of engineering recently.  In most of such attempts, atoms or ions are isolated in an ultra-high vacuum (UHV) environment in the range of $10^{-8}$--$10^{-10}$~Pa, hence methods of generating and manipulating atoms and ions without degrading the UHV environment are of essential importance in terms of the stability and reproducibility of cold-atom experiments.  
Generation of atoms and ions, in this respect, is likely the only procedure that may drastically affect the environment.  Traditional method of atom generation is an oven which utilizes resistive heating of a wire to generate atomic vapors from metals or compounds~\cite{Schioppo2012-cf}.  This atom oven or dispenser is relatively easy to implement but results in excess amount of atomic vapor and non-negligible heat generation particularly for materials possessing low saturated vapor pressures, each being one of the major obstacles for achieving long trapping lifetime and long-term stability.   

\begin{figure}[b]
\includegraphics[width=8.6cm]{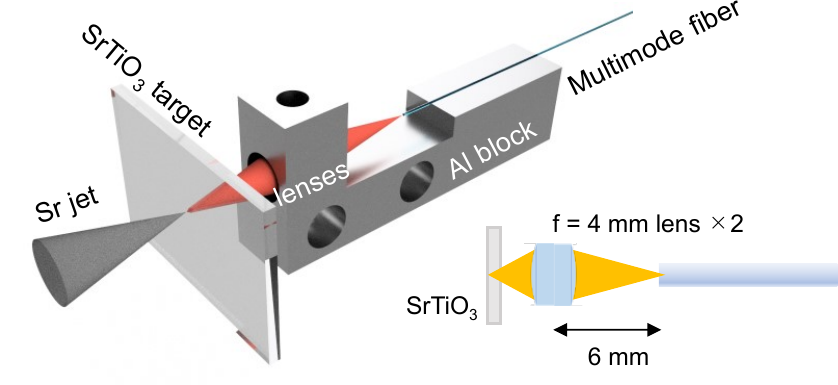}
\caption{\label{fig1} Schematic illustration of our atom source. A multimode fiber and a focusing lens pair are fixed to an aluminum block. A lens pair consists of two lenses with their focal length of 4~mm and placed $\sim$6~mm away from the cleaved facet of the multimode fiber. 
A 0.5~mm-thick SrTiO$_3$ plate is placed at the focal point to be ablated by the pulsed laser.}
\end{figure}

Laser ablation or laser-induced heating as an alternative method has been investigated for a couple of decades and is successful for atomic species such as Sr and Yb for neutral-atom experiments~\cite{Yasuda2017-ah, Hsu2022-wm}, and $^{40}$Ca$^+$~\cite{Shao2018-ak}, $^9$Be$^+$~\cite{Sameed2020-ac}, $^{88}$Sr$^+$~\cite{Leibrandt2007-au}, $^{174}$Yb$^+$~\cite{Vrijsen2019-xr} and other ions~\cite{Piotrowski2020-kz,Olmschenk2017-mf} for ion-trap experiments. It is even valid for generating low-energy electrons combined with photoionization~\cite{Osada2022-hy}.  Pulsed laser ablation has advantages that the heat generation is local and nanosecond-short if a focused, nanosecond laser pulse is used, so that it is even compatible with the ion trapping in a cryogenic environment~\cite{Niemann2019-yk,Dubielzig2021-qh}
getting increasingly important in persuit of ion-trap quantum computer.  On top of that, loading of only a single ion can be executed with 82~\% probability as shown in our previous work~\cite{Osada2022-oh}.  One drawback is that one should shine an intense pulsed laser on a target in a vacuum setup by adding optics in existing laser-cooling optics densely surrounding the vacuum shroud, thence the vacuum setup needs additional care and added footprints on the optical table cannot be ignored.

In this paper, we solve this issue by simply shining nanosecond YAG-laser pulses delivered via a multimode optical fiber and focused by a miniature lens system, and constructed a compact, fiber-pigtailed pulsed-laser-ablation-based atom source.  Our device consists of a multimode optical fiber, a focusing lens and an ablation target, where the fiber and lens are mounted on the same aluminum block. The device footprint is about a few-milimeter wide, several-milimeter tall and less than 20-mm long, except the fiber pigtail.  Our device is capable of generating strontium atoms which is confirmed by observing the resonance fluorescence, by ablating a SrTiO$_3$ target.  Furthermore, by analyzing the behavior of the resonance fluorescence, one can infer device characteristics such as velocity, temperature and the number of atoms of the generated strontium atomic jet. The atomic source realized in this work is useful not only for cold-atom, especially the ion-trap, experiments but also for other material-science experiments demanding a compact, modular, cryo-compatible, pulsed beam sources.


As a gross structure, our atom source is constituted by an optical fiber and commercial miniature lenses, both mounted on a single aluminum block, with an ablation target placed at the focal point of the outcoupled laser beam.
In this paper, the target material of the laser ablation is SrTiO$_3$ from which the strontium atoms are expected to be shot out.  A schematic illustration of our device is shown in Fig.~\ref{fig1}.  In the scope of using this device in an UHV environment, all components are fixed either by just screwing or otherwise glued with TorrSeal that allows for the use in $10^{-8}$~Pa or less. 

In previous study~\cite{Osada2022-oh}, the pulsed laser ablation of SrTiO$_3$ is revealed to work with a nanosecond, pulsed Nd:YAG laser of 4~J/cm$^2$ fluence which can be realized by pulse energy of less than 200~$\mu$J and moderate focusing of the beam.  An optical fiber used for our device should not be damaged by irradiation of such laser pulses at first place.  Laser-induced damage threshold of fused silica, a most widely-used material for optical fibers, is studied in Cao~\textit{et al.}~\cite{Cao2018-yc} and Nieto \textit{et al.}~\cite{Nieto2015-hy} which report the values exceeding 800~J/cm$^2$.  This value is, fortunately, well above that of the SrTiO$_3$, so that there securely exists some condition that the SrTiO$_3$ is ablated but the optical fiber is not, without the elaboration on tight focusing of the ablation laser on the target.  We confirmed that a 1064~nm-wavelength, 10~ns-pulsed laser of pulse energy below 300~$\mu$J can be guided over 6\,000 times without degradation of the coupling efficiency around 80~\% through multimode fibers with core diamters larger than 50~{$\mu$}m, while a single-mode fiber and a multimode fiber with 25~{$\mu$}m core get gradually damaged.  No variation of the pulse width is detected after the pulse is delivered through the fiber under such conditions.  From consideration on the long-term durability and expected pulse fluence, we adopt a multimode fiber with 105~$\mu$m-core here throughout this paper.  

The fiber is cleaved and glued to the aluminum block (see Fig.~\ref{fig1}), on which two 2~mm-diameter, plano-convex lenses of focal length being 4~mm are mounted 6~mm away from the cleaved edge of the fiber, see inset of Fig.~\ref{fig1}. With this configuration, transmitted laser pulse is focused at $\sim$3~mm off the lens where the ablation target is placed, though the lens system after the fiber output should act as a reduction optics of its magnification being 0.5, spherical aberration gives the extended spot size of around $200$~$\mu$m.  Crystalline SrTiO$_3$ is a highly transparent material at the wavelength of 1064~nm, and interestingly the pulsed laser ablation can take place not only on the incident surface but also on the output surface.  Thanks to the short focal length that is comparable to the thickness of the SrTiO$_3$ of 0.5~mm, we can choose which surface to ablate.  Here we placed the SrTiO$_3$ sample so that only the output surface is ablated, a configuration that minimizes the unwanted material deposition onto the lens and facilitates the construction of particle filters, e.g. a metallic mesh or pinhole, for eliminating the unwanted charged particles generated in the ablation.  

\begin{figure}[t]
\includegraphics[width=8.6cm]{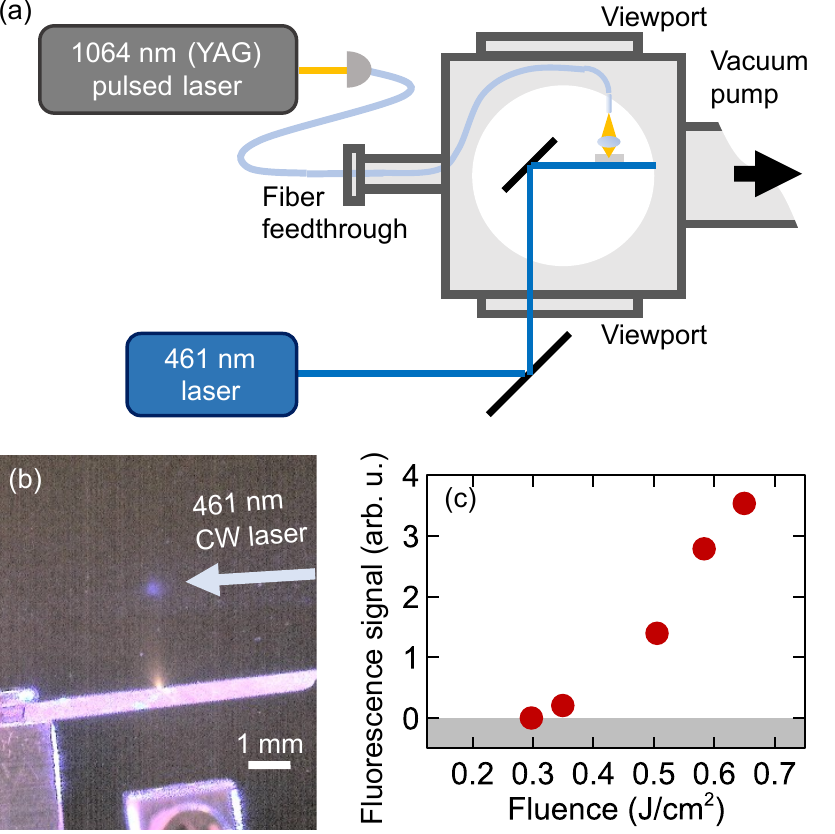}
\caption{\label{fig2} Demonstration of strontium generation with our device.  (a) Experimental setup. A 1064~nm-wavelength, nanosecond pulsed laser is coupled to a 105~$\mu$m core fiber and the fiber-based atom source is placed in a vacuum shroud through a fiber feedthrough.  A continuous-wave 461~nm-wavelength laser is led into the shroud via one of viewports for driving $^1S_0$-$^1P_1$ transition of generated neutral strontium atoms.  Then the fluorescence is observed by an imaging system with numerical aperture $\sim 0.05$ through another viewport.  (b) Observed laser ablation and subsequent fluorescence of strontium atoms. (c) Integrated counts of fluorescence for various values of pulse fluence.  Eight ablation events are added up and fluorescence signals are integrated in each data point.  The background level, corresponding to the fluorescence count with $~\sim$0.3~J/cm$^2$ is subtracted.  }
\end{figure}


\begin{figure}[t]
\includegraphics[width=8.6cm]{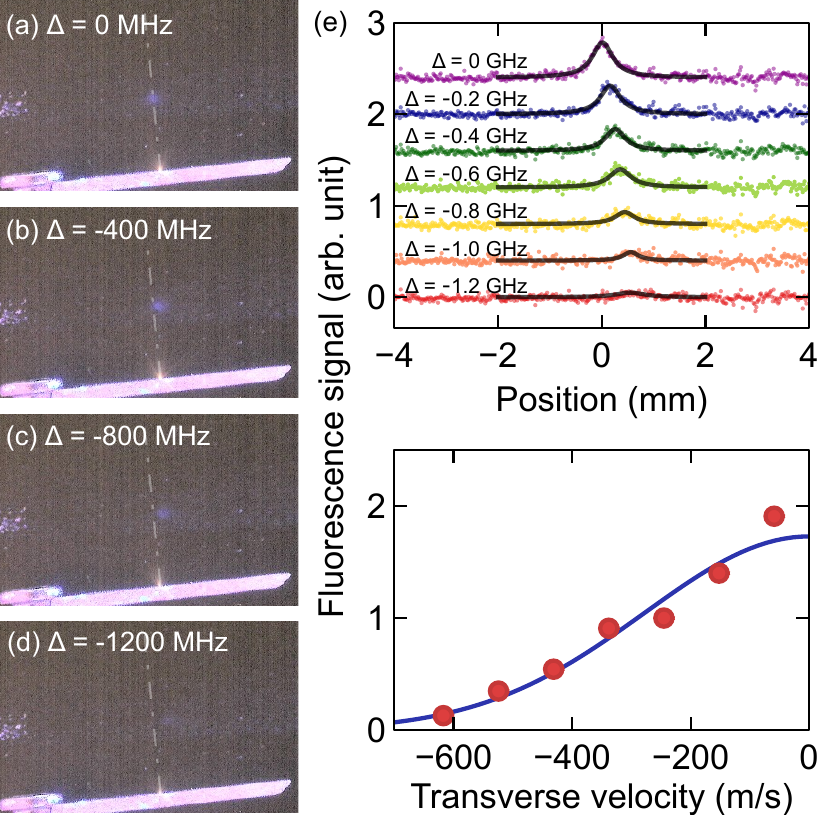}
\caption{\label{fig3} Characterization of jet of atoms generated by pulsed laser ablation with our device. (a-d) Fluorescence images for various detunings $\Delta = f_\mathrm{laser}-f_0$ indicated in each image. Dash-dotted lines normal to the SrTiO$_3$ surface are provided as eye-guide. (e) Transverse distributions of fluorescence signals for $\Delta =$~0~GHz (purple), $-0.2$~GHz (blue), $-0.4$~GHz (green), $-0.6$~GHz (light green), $-0.8$~GHz (yellow), $-1.0$~GHz (orange) and $-1.2$~GHz (red) from top to bottom. Black solid curves are Lorentzian fittings. (f) Transverse-velocity distribution. Red datapoints represent integrated fluorescence signals where detunings are reinterpreted as velocities. Blue curve is a Gaussian fitting from which transverse temperature of the jet of atoms is estimated to be 800~K.}
\end{figure}

To demonstrate that our device can actually generate strontium atoms, we constructed an experimental setup depicted in Fig.~\ref{fig2}(a).  A 1064~nm-wavelength, nanosecond pulsed laser of pulse energy $< 300$~$\mu$J is coupled to a 105~$\mu$m core fiber and introduced into a vacuum shroud through a fiber feedthrough. The fiber-based atom source is placed in a vacuum shroud with a pressure of $\sim 4 \times 10^{-4}$~Pa.  We note again that the whole system of fiber-based ablation is made of a metallic block, screw and UHV-compatible glue so that it can be securely installed into UHV environment and baked at 400~K.  A continuous-wave 461~nm-wavelength laser is led into the shroud via one of viewports for driving $^1S_0$-$^1P_1$ transition of generated neutral strontium atoms. The laser power is 12~mW with a beam diameter of 0.14~mm, which yields the laser intensity well above the saturation intensity $I_\mathrm{sat} = 43$~mW/cm$^2$. The fluorescence is observed by an imaging system with numerical aperture $\sim 0.05$ through another viewport which is not shown in Fig.~\ref{fig2}(a).

Figure~\ref{fig2}(b) displays a typical optical micrograph that signaling a successful generation of strontium atoms with our device.   The ablation laser is fed from the bottom and focused by lenses mounted in an aluminum block seen below. The ablation laser then passes through the SrTiO$_3$ substrate and by aligning the output surface at the very focus of the ablation laser, the output surface is ablated, as the white, bright flash can be seen.  The blue fluorescence can be observed at a point where 461~nm-wavelength laser passes, where the laser frequency is tuned at $650\,503.8$~GHz that is almost on resonance on the $^1S_0$-$^1P_1$ transition of the neutral strontium atom.  The image acquisition time is set to be 4~s and the ablation pulse is fired at 2~Hz repetition, therefore 8 images are accumulated to result in a single fluorescence image here.


As a device characterization, we first checked the threshold behavior of the laser ablation by varying the pulse energy and hence the pulse fluence.  The observed strontium fluorescence signals for various pulse fluence are plotted in Fig.~\ref{fig2}(c).  
The background signal level (top of the gray hatched region)
is subtracted for better visibility.  The fluorescence becomes visible with the fluence around 0.4~J/cm$^2$ in our setup.  

Next, we characterize the jet of Sr atoms generated in our experiment, especially in terms of the transverse temperature and longitudinal velocity. Here we define the longitudinal direction to be the most probable direction in which atoms are ejected, and the transverse direction to be perpendicular to the longitudinal one.  As the nanosecond pulsed laser ablation principally rely on the light absorption and heating at the target surface, the longitudinal direction can be safely assumed to coincide the normal direction of the target surface. We assume in the following analysis that atoms are ejected from a point where ablation takes place and flies ballistically, the latter of which is justified with the low vacuum pressure.   In order to investigate above quantities, we tuned the laser frequency $f_\mathrm{laser}$ red-detuned from the center frequency of the Doppler-broadened resonance $f_0 = 650\,503.8$~GHz and examine the behavior of the fluorescence spot with respect to the detuning $\Delta = f_\mathrm{laser} - f_0$.  The results are shown in Fig.~\ref{fig3}(a-d), where the shifts of the fluorescence spots can be seen in the microscope images. In the top panel of Fig.\ref{fig3}(e), transverse distributions of the fluorescence signals (colored dots) and results of Lorentzian fitting (dark gray lines) are shown for detunings indicated on the right, with corresponding vertical offsets for visibility. Horizontal axis represents the position along the transverse direction where origin is set to the peak of the fluorescence signal for $\Delta = 0$~GHz.   These spatial shifts are attributed to the Doppler effect with transverse velocity $v_\perp$ of atoms, that is, the velocity perpendicular to the normal direction to which the most of ablation plume is shuttled. The detuning $\Delta$ can be directly converted into $v_\perp$ through $\Delta = f_\mathrm{laser} v_\perp /c$ with $c$ being the speed of light, and we can plot the velocity distribution by collecting the fluorescence counts, as displayed by red points in the bottom panel of Fig.~\ref{fig3}(f).  By assuming a thermal distribution, we can fit the data with a Gaussian distribution and estimate the transverse temperature of the atom jet to be 800~K.

Furthermore, the velocity deduced from $\Delta$ and the shift of transverse position in Fig.~\ref{fig3}(e) are combined to allow us to guess the time duration required for atoms to arrive at the detected position.  This time duration and longitudinal position of the detected atoms yields a value of longitudinal velocity $v_\parallel$, a typical value of the velocity along the atom jet.  The time required to reach the detected position is about 1.1~$\mu$s, with which 2.4~mm is traveled perpendicular to the surface of the SrTiO$_3$.  From these values, we can estimate the longitudinal velocity to be $v_\parallel = 2.3 \times 10^3$~m/s.

As a further characterization, we roughly estimate the number of atoms detected by a single ablation event as follows.  We integrate the fluorescence signals as shown previously in Fig.~\ref{fig2}(b) to deduce the number of photons emitted by strontium atoms through a calibration of the sensitivity of the microscope system with a weak coherent laser.  Since the natural linewidth of the $^1S_0$-$^1P_1$ transition of neutral Sr is $2\pi \times 33$~MHz and is power-broadened with the intense laser, the lifetime of the irradiated Sr atoms is 76~ps. Such atoms passing through the beam diameter of 480~$\mu$m with their velocity $v_\parallel = 2.3 \times 10^3$~m/s can experience 2\,800 transition cycles.  The last factor is the collection efficiency $\sim 6 \times 10^{-4}$ of the microscope system which is simply calculated from the numerical aperture of about 0.05.  Combining these factors together, the number of detected strontium atoms per single ablation pulse is estimated to be $2 \times 10^5$ when the fluence is about 0.64~J/cm$^2$ and $\Delta = 0$~GHz.  


In conclusion, we designed, demonstrated, and characterized an fiber-based atom source based on pulsed laser ablation.  By using commercially available miniature lens system for focusing nanosecond pulsed laser of up to 225~$\mu$J delivered through a multimode fiber of 105~$\mu$m core, we successfully ablated a SrTiO$_3$ target and generated a jet of neutral strontium atoms, while our method can be applied to other transparent ablation targets containing materials under concern.  Our device endured 6\,000 cycles of ablation event without noticeable damage on the fiber facets and lenses.  The generated strontium beam exhibited the transverse temperature of 800~K and longitudinal velocity of 2\,300~m/s, which are typical of pulsed-laser-ablation-based atom source. The number of atoms detected in a single ablation pulse was estimated to be $2 \times 10^5$ that is sufficient not only for ion-trap experiments but possibly for spectroscopic and interferometric experiments using neutral atoms. Our device provides a compact, cryo-compatible fiber-pigtailed atom source with small device footprints in the vacuum shroud and reduced complexity of vacuum systems to further promote the developments of cold-atom experiments. It may also find interesting applications in atomic and molecular sciences requiring cryo-compatible, small atom and/or ion sources combined with particle filters.

This work was supported by JST PRESTO (Grant No. JPMJPR1904) and JST Moonshot R\&D (Grant No. JPMJMS2063-5-2) programs.

\bibliography{main}

\end{document}